\documentclass[prl,twocolumn,showpacs,superscriptaddress,preprintnumbers]{revtex4-2}
\usepackage{amsmath,amssymb,amsfonts,float,graphics,epsfig,epstopdf,color,verbatim,tabularx,bm,multirow,appendix,ulem}
\usepackage{tikz}
\usepackage{amsmath,amssymb,graphicx}
\usepackage{wasysym}
\usepackage[utf8]{inputenc}
\usepackage[T1]{fontenc}
\usepackage{xcolor}
\usepackage{dsfont}
\usepackage{textcomp}
\usepackage{bm}

\usepackage{graphicx}%
\usepackage{xcolor}
\usepackage{dcolumn}
\usepackage{bm}
\usepackage{xspace}
\usepackage{time}
\usepackage{booktabs}
\usepackage{multirow}
\usepackage{makecell}

\pdfoutput=1
\usepackage{color}
\definecolor{LinkColor}{rgb}{0.256,0.439,0.588}
\usepackage{hyperref}
\hypersetup{
colorlinks=true,
citecolor=LinkColor,
linkcolor=LinkColor,
urlcolor=LinkColor
}

\newcommand{\newsect}[1]{\noindent \textit{\textcolor{blue}{#1.-}}}

\begin{document}

\title{Thermodynamics of Shastry-Sutherland Model under Magnetic Field}
\author{Menghan Song}
\affiliation{Department of Physics and HK Institute of Quantum Science \& Technology, The University of Hong Kong, Pokfulam Road,  Hong Kong SAR, China}
\affiliation{State Key Laboratory of Optical Quantum Materials, The University of Hong Kong, Pokfulam Road,  Hong Kong SAR, China}
\date{January 2026}

\author{Chengkang Zhou}
\affiliation{Department of Physics and HK Institute of Quantum Science \& Technology, The University of Hong Kong, Pokfulam Road,  Hong Kong SAR, China}
\affiliation{State Key Laboratory of Optical Quantum Materials, The University of Hong Kong, Pokfulam Road,  Hong Kong SAR, China}
\date{January 2026}

\author{Cheng Huang}
\affiliation{Department of Physics and HK Institute of Quantum Science \& Technology, The University of Hong Kong, Pokfulam Road,  Hong Kong SAR, China}
\affiliation{State Key Laboratory of Optical Quantum Materials, The University of Hong Kong, Pokfulam Road,  Hong Kong SAR, China}
\date{January 2026}

\author{Zi Yang Meng}
\email{zymeng@hku.hk}
\affiliation{Department of Physics and HK Institute of Quantum Science \& Technology, The University of Hong Kong, Pokfulam Road,  Hong Kong SAR, China}
\affiliation{State Key Laboratory of Optical Quantum Materials, The University of Hong Kong, Pokfulam Road,  Hong Kong SAR, China}
\date{January 2026}

\begin{abstract}
Motivated by the recent experimental discovery of the $T$-linear specific heat in pressurized and magnetized Shastry-Sutherland Mott insulator SrCu$_2$(BO$_3$)$_2$, we perform the state-of-the-art thermal tensor-network computation on the Shastry-Sutherland model under a magnetic field. Our simulation results suggest the existence of a symmetric intermediate phase with $T$-linear specific heat at low temperature, occupying a large parameter space between the plaquette-singlet and antiferromagnetic phases at low fields, and separates other symmetry-breaking phases at high fields before full polarization. Such an unexpected novel state bears an astonishing similarity to the experimental findings in the material. It opens the door to further investigations of the possible liberation of deconfined magnetized Dirac spinons by the competing interactions in this highly frustrated quantum magnet model, and by the combined effects of magnetic field and pressure in the the associated Shastry-Sutherland Mott insulator SrCu$_2$(BO$_3$)$_2$.
\end{abstract}
\maketitle

\newsect{Introduction}
The Shastry-Sutherland (SS) model~\cite{Shastry1981Exact}, originally introduced as a rare example of a two-dimensional quantum antiferromagnet with an exact dimer-singlet ground state, has become a cornerstone in the study of frustrated magnetism. Its physical realization in the layered compound $\text{SrCu}_2(\text{BO}_3)_2$~\cite{Kageyama1999Exact, Miyahara1999Exact} has provided a unique experimental platform to explore a wide variety of quantum phases and transitions. At ambient pressure, $\text{SrCu}_2(\text{BO}_3)_2$ is in the exact dimer-singlet (DS) phase. When increasing the pressure, the ratio of the inter-dimer ($J$) coupling to the intra-dimer coupling $J^\prime$ enhances, triggering transitions into a plaquette-singlet (PS) phase~\cite{Zayed20174-spin,Guo2020QuantumPhases,Jiménez2021analogue,shiDiscovery2022,Guo2025Deconfined,Cui2023Proximate} and eventually an antiferromagnetic (AF) phase at higher pressures~\cite{Guo2020QuantumPhases,Cui2023Proximate,Guo2025Deconfined,Zayed20174-spin,Jiménez2021analogue}. The nature of the transition between the PS and AF phases has attracted significant attention, as it stands for a potential candidate for a deconfined quantum critical point (DQCP)~\cite{Senthil2004Deconfined, Senthil2004Quantum,sandvikEvidence2007,maDynamical2018}, where the conventional Landau-Ginzburg-Wilson paradigm of phase transition may be insufficient to describe the situation. However, recent advanced high-pressure experiments suggest a first-order transition from the observation that the transition temperatures of the PS and AF phases meet at a bi-critical point at a finite temperature and pressure~\cite{Guo2025Deconfined}.

\begin{figure}[htp!]
	\includegraphics[width=\columnwidth]{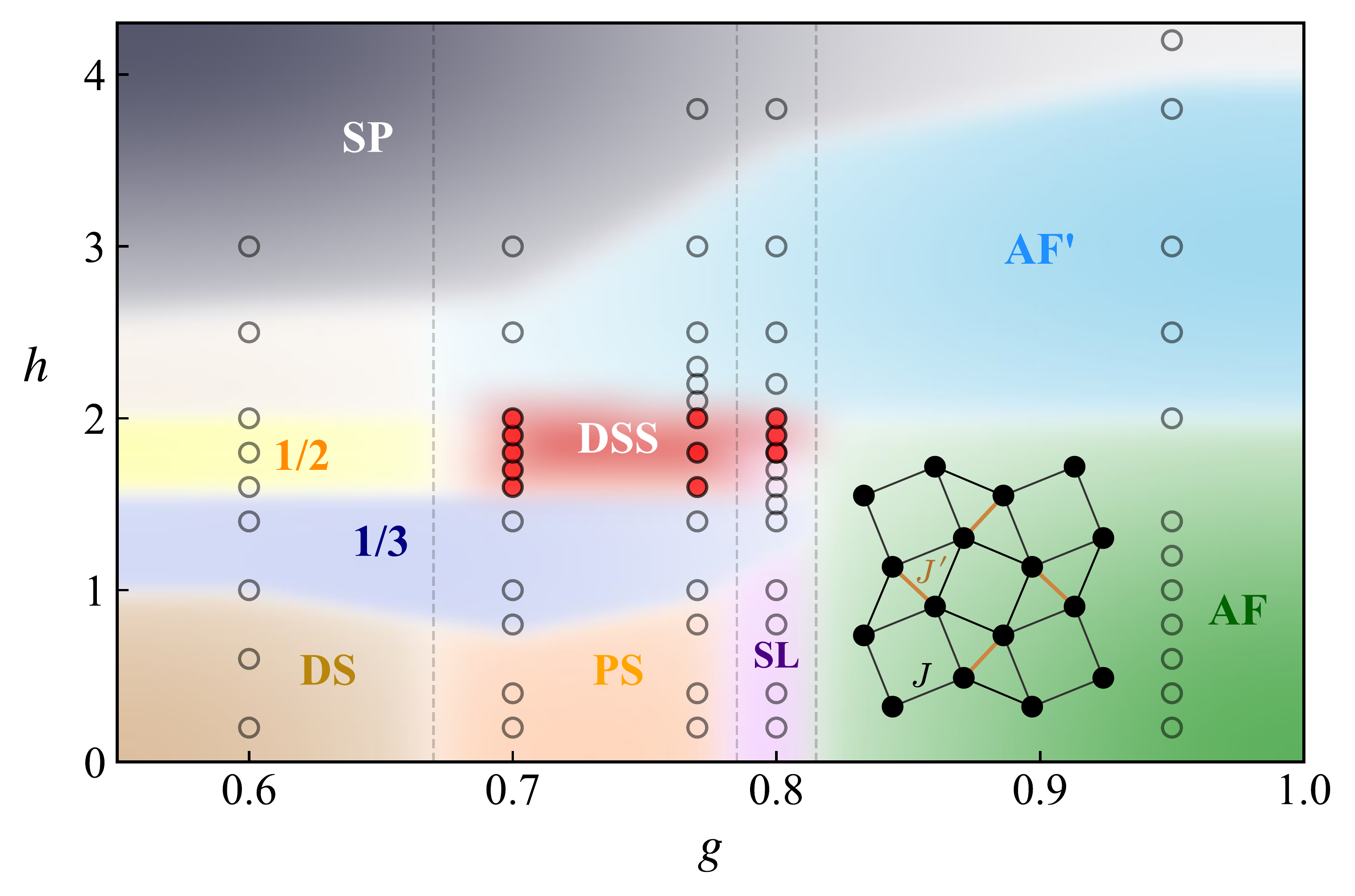}
	\caption{\textbf{{Phase diagram of the SS model under a magnetic field.}} The previously known zero-field phase boundaries are shown as dashed lines. The dots indicate parameter points studied via XTRG simulations, with the red dots highlighting regions where a linear temperature dependence of the specific heat is observed, serving as a signature for the Dirac spinon state (DSS). The remaining acronyms denote the spin-polarized (SP), dimer singlet (DS), plaquette singlet (PS), spin liquid (SL), antiferromagnetic (AF), and field-induced antiferromagnetic (AF$'$) phases. Vertical dashed lines indicate the zero-field phase boundaries obtained by previous works~\cite{Guo2020QuantumPhases,Corboz2013Tensor,Koga2000Quantum,Yang2022Quantum,Viteritti2025Transformer,maityEvidence2025,Corboz2025quantum}. The inset shows the lattice of the SS model.}
	\label{fig:fig1}
\end{figure}

The numerical investigation of the SS model's phase diagram has been a long-standing challenge due to the high degree of frustration. The model (Eq.~\eqref{eq:hamiltonian}) inside the DS phase and the model with ferromagnetic transverse and antiferromagnetic interactions have been simulated with the quantum Monte Carlo method~\cite{mengPhases2008,Wessel2018Thermodynamic,Wietek2019Thermodynamic}. However, the original model at $h=0$ close to and beyond the DS and PS transition, has to be solved with other methods due to the sign-problem. For the tuning parameter $g\equiv J/J^\prime$, early studies with exact diagonalization and tensor-network methods have identified the existence of DS, PS, and AF phases at small, intermediate, and large $g$, respectively~\cite{Corboz2013Tensor,Koga2000Quantum,Lee2019Signatures,Nakano2018Third}. More recently, advanced numerical techniques have been employed to resolve the controversial region between the PS and AF phases. Density matrix renormalization group (DMRG) simulations on cylinders suggested the existence of a narrow gapless quantum spin liquid (SL) phase intervening between the PS and AF orders~\cite{Yang2022Quantum,maityEvidence2025}. This was further supported by the most recent works by neural-network quantum states (NQS) using Vision Transformers~\cite{Viteritti2025Transformer} and high-precision infinite projected entangled-pair states (iPEPS) simulations~\cite{Corboz2025quantum}, where the SL phase is found in a narrow range of $0.78 \lesssim g \lesssim 0.82$.

Experimentally, the behavior of $\text{SrCu}_2(\text{BO}_3)_2$ under a magnetic field has revealed even more exotic physics. While the zero-field PS-AF transition in the real material is found to be first-order~\cite{Guo2025Deconfined}, recent high-pressure calorimetry measurements have uncovered novel gapless critical behavior emerging in this insulator under magnetic field~\cite{Guo_T-linear}. Specifically, in the vicinity of the PS-AF transition, the application of a magnetic field leads to a $T$-linear dependence in the specific heat $C$, with $C/T$ remaining constant as $T \to 0$. This is a hallmark of metallic behavior in a well-defined charge Mott insulator~\cite{Guo2025Deconfined,Guo_T-linear}. It hints at a Dirac quantum spin liquid under a magnetic field associated with a Fermi surface of fractionalized spinon excitations~\cite{shimizuSpin2003,leeU1Gauge2005,motrunichOrbit2006,Ran2007Projected,leeAmperean2007,yamashitaThermodynamic2008,yamashitaGapless2011,Niravkumar2019Magnetic,Wu2025SpinDynamics} and hence the phase is dubbed the Dirac spinon state (DSS).

Facing the rapid experimental progress, the thermodynamics of the SS model becomes even less well understood theoretically, especially under maganetic field. So far, studies have been limited to the DS phase~\cite{Wessel2018Thermodynamic,Wietek2019Thermodynamic}, and the effects of a magnetic field remain entirely uninvestigated. The recently discovered intermediate DSS phase in the pure SS model raises fundamental questions about how this phase evolves and responds to the magnetic field. In this Letter, we address this question by employing the exponential thermal tensor network (XTRG) technique~\cite{Chen2018Exponential} to systematically investigate the finite-temperature thermodynamic properties of the Shastry-Sutherland model under an external magnetic field. By performing simulations on quasi-one-dimensional cylinder geometries, we calculate various physical observables at finite temperatures and infer a semi-quantitative zero-temperature phase diagram in the $h$-$g$ parameter space, as shown in Fig.~\ref{fig:fig1}. 

Drawing inspiration from high-pressure experiments~\cite{Guo2025Deconfined,Guo_T-linear}, our numerical simulations provide indications consistent with the presence of phases observed in experimental $h$-$g$ phase diagrams, including the presence of spin-polarized (SP), dimer singlet (DS), plaquette singlet (PS), antiferromagnetic (AF), and field-induced antiferromagnetic (AF$'$) phases. Notably, within the constraints of our investigated system size and bond dimension, we observe a $T$-linear dependence of the specific heat $C$ at intermediate magnetic field strengths within the PS phase and near the PS-AF phase boundary for several values of $g$. In line with the experimental result, we dub this insulating state with metallic behavior as Dirac spinon state (DSS) in our proposed phase diagram (Fig.~\ref{fig:fig1}), offers theoretical insights that align with recent experimental observations in $\text{SrCu}_2(\text{BO}_3)_2$ and contributes to understanding the stability and characteristic signatures of the Dirac spin liquid phase in the presence of magnetic fields.

\begin{figure}[htp!]
	\includegraphics[width=\columnwidth]{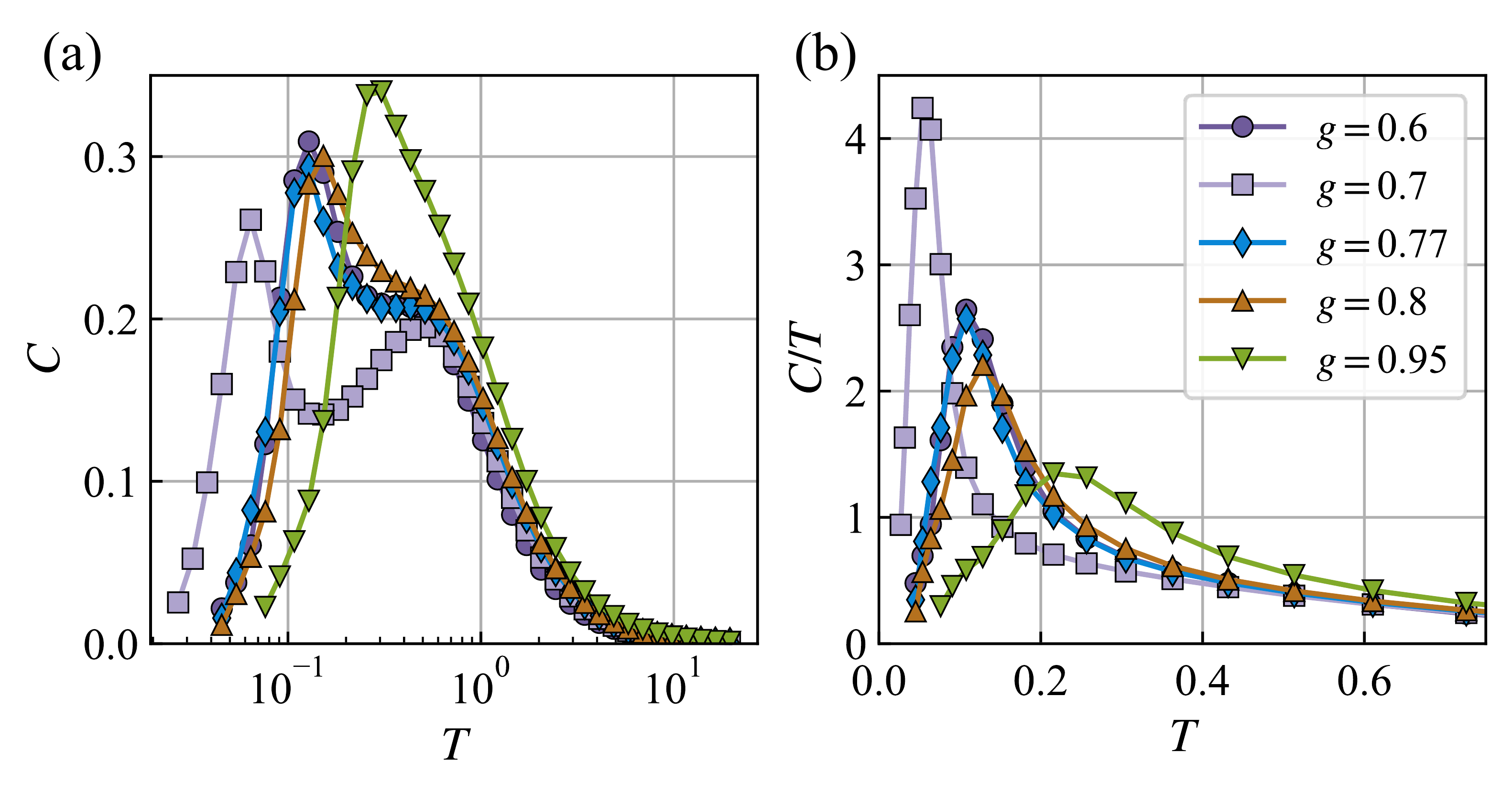}
	\caption{\textbf{Specific heat in the SS model at zero magnetic field.} Temperature dependence of (a) specific heat $C$ and (b) specific heat divided by temperature $C/T$ for different coupling strengths. $g=0.6$ is inside the DS phase, $g=0.7$ is inside the PS phase, $g=0.77,0.8$ are close to the PS-AF transition and $g=0.95$ is inside the AF phase.} 
	\label{fig:fig2}
\end{figure}

\begin{figure*}[htp!]
	\centering
	\includegraphics[width=\textwidth]{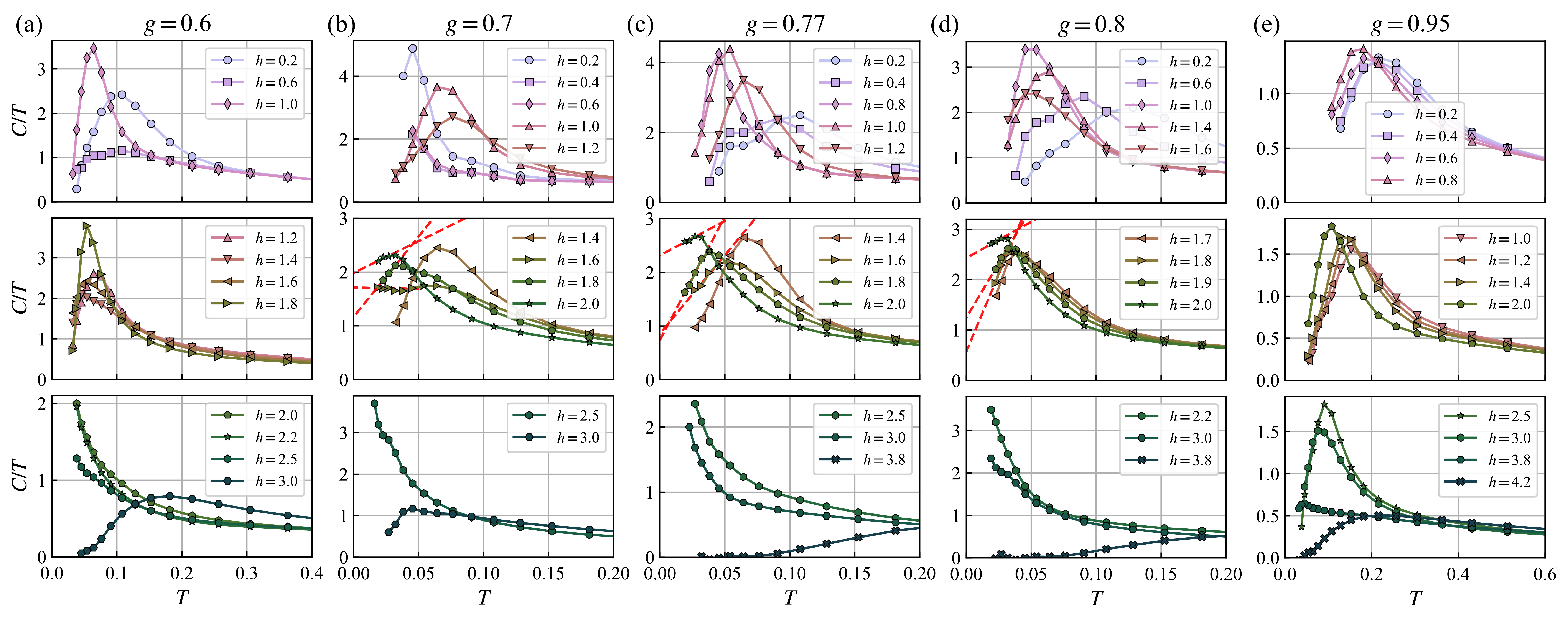}
	\caption{\textbf{Specific specific heat divided by temperature, $  C/T  $, as a function of temperature $  T  $ for various selected magnetic fields $  h  $, at different coupling constants $g$ for different phases.} Panels (a)-(e) display: (a) DS phase ($  g=0.6  $), (b) PS phase ($  g=0.7  $), (c) PS--SL phase boundary ($  g=0.77  $), (d) SL phase ($  g=0.8  $), and (e) AF phase ($  g=0.95  $). The red dashed line represents a linear extrapolation of the last three $C/T$ data points based on $C/T = a_0 +a_1T$, which intercepts the vertical axis at a finite value $a_0$. This finite intercept implies a linear dependence of the $C$ against $T$. Noisy data points at very low temperatures were omitted because of truncation errors and inaccuracies in the numerical derivative used to obtain the specific heat.}
	\label{fig:Cv_over_T}
\end{figure*}

\newsect{Models and Methods}
The SS model under the magnetic field is described by the Hamiltonian:
\begin{equation}
H = J \sum_{\langle i,j \rangle} \mathbf{S}_i \cdot \mathbf{S}_j + J' \sum_{\langle\langle i,j \rangle\rangle} \mathbf{S}_i \cdot \mathbf{S}_j - h \sum_i S_i^z
\label{eq:hamiltonian}
\end{equation}
where $\mathbf{S}_i$ are spin-1/2 operators on site $i$. The first term represents the antiferromagnetic inter-dimer coupling $J$ between nearest-neighbor spins on a square lattice, while the second term describes the intra-dimer coupling $J'$, as denoted in the inset of Fig.~\ref{fig:fig1}. We set $J'=1$ as the energy unit throughout this paper. The third term introduces a Zeeman coupling to an external magnetic field $h$ along the $z$-direction. The ratio $g\equiv J/J^{\prime}$ serves as the primary tuning parameter for the zero-field phase diagram, which has been determined based on previous numerical simulations with ED and tensor-network methods over the years~\cite{Guo2020QuantumPhases,Corboz2013Tensor,Koga2000Quantum,Yang2022Quantum,Viteritti2025Transformer,maityEvidence2025,Corboz2025quantum}. For $g\lesssim0.67$, the ground state is the DS phase. The PS phase is identified in the range $0.67\lesssim g\lesssim 0.78$, while the AF phase occurs for $g\gtrsim 0.82$. Between the PS and AF phases, a narrow SL phase is found.

We employ the exponential thermal tensor network (XTRG) algorithm~\cite{Chen2018Exponential} to investigate the thermodynamic properties of the SS model under a magnetic field. A detailed description of the XTRG method is provided in the End Matter (EM) for interested readers. The key advantage of XTRG is its exponential evolution scheme, which doubles the inverse temperature at each step, significantly reducing the number of imaginary-time evolution and truncation operations required. This efficiency allows us to obtain accurate thermodynamic data down to very low temperatures, such as $T \approx 0.01J'$ in our work. Since in the material, $J' \sim 80$K at ambient pressure~\cite{Guo2020QuantumPhases}, our simulation temperature scale is comparable with the scale DSS is seen in $\text{SrCu}_2(\text{BO}_3)_2$, i.e. $\sim 1$K. 


For efficiency, we implement symmetries into matrix product operator (MPO) tensors based on the TensorKit framework~\cite{devos2025tensorkit}. We implement the $SU(2)$ symmetry for the simulation at zero fields and keep $D=2000$ bond states. With a non-zero Zeeman field, we implement the $U(1)$ symmetry with $D = 800$ bond states kept, maintaining a maximum truncation error at approximately $10^{-2}$. Our simulations are performed on a Y-cylinder geometry, meaning the system has periodic boundary conditions along the $y$ direction and open boundary conditions along the $x$ direction. The simulation size is set to $3 \times 12 \times 4$, where $3$ and $12$ denote the number of unit cells along $y$ and $x$ directions, respectively, and $4$ represents the number of sites in a unit cell. The lattice geometry is exemplified in Fig.~\ref{fig:lattice} in the EM.

\begin{figure*}[htp!]
	\centering
	\includegraphics[width=\textwidth]{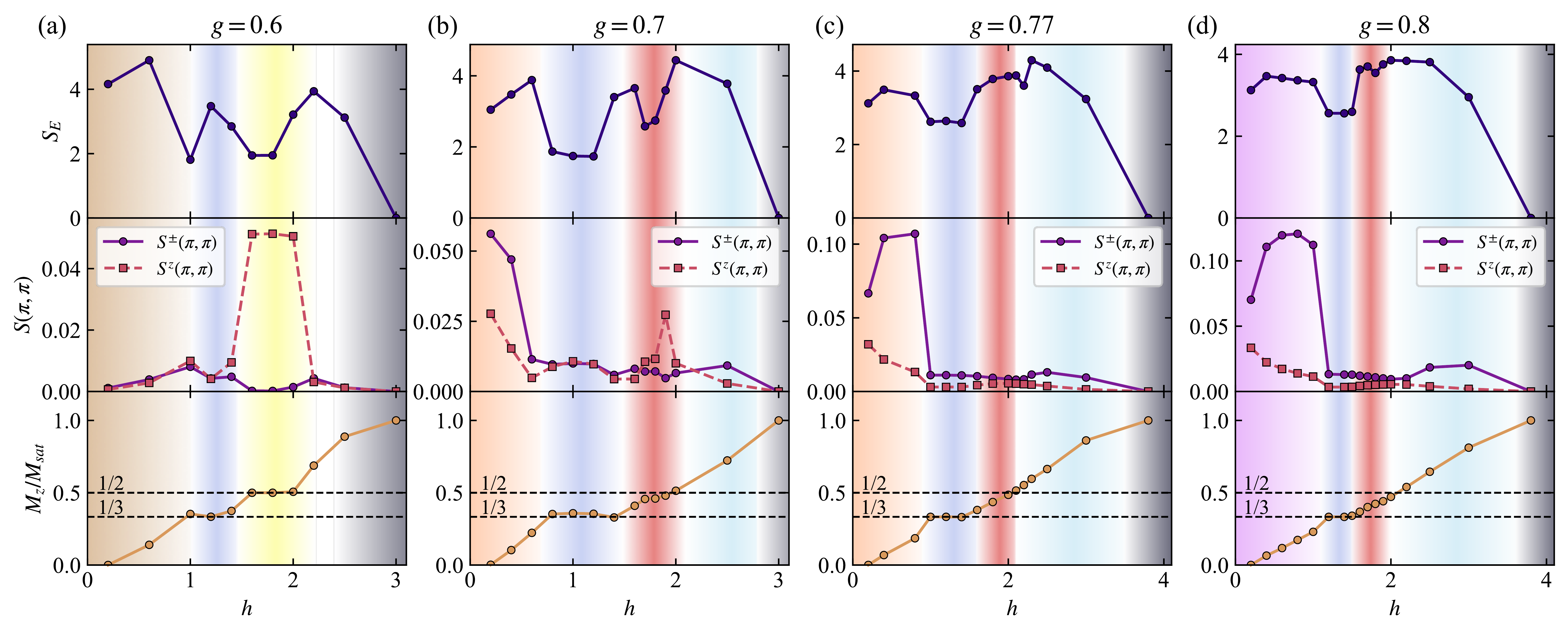}
	\caption{\textbf{Physical observables (entanglement entropy, magnetic structure factor, and magnetization) computed via XTRG across coupling constants $g$:} (top) half-system thermal entanglement entropy $S_E$, (middle) spin structure factor at $\mathbf{q}=(\pi,\pi)$, $S^{\pm(z)}(\pi,\pi) 
= \frac{1}{N} \sum_{i,j=1}^{N} e^{i\mathbf{q}\cdot(\mathbf{r}_i-\mathbf{r}_j)} 
\langle {S}^{\pm(z)}_i \cdot {S}^{\pm(z)}_j \rangle$, (bottom) normalized magnetization, $M_z/M_{sat}= \frac{2}{N} \left| \sum_{i=1}^N \langle S_i^z \rangle \right|$. Observables are measured at temperature $T\approx0.01$. All quantities are shown as functions of magnetic field $h$, with color scheme indicating phase regimes consistent with Fig.~\ref{fig:fig1}. The dashed lines in the bottom panel indicate the 1/3 and 1/2 magnetic plateaus.}
	\label{fig:fig4}
\end{figure*}

\newsect{Zero field results}
Before analyzing the effect of a magnetic field, we first examine the specific heat at zero field, $h=0$. Fig.~\ref{fig:fig2}(a) presents $C$ at some representative $g$ values. Deep in the DS phase at $g=0.6$, we observe the characteristic double-peak structure: a sharp low-temperature peak at $T \sim 0.1$ and a broader high-temperature shoulder at $T \sim 0.5$, consistent with prior numerical studies~\cite{Wessel2018Thermodynamic,Wietek2019Thermodynamic} and experiments~\cite{Guo2025Deconfined,Guo_T-linear}. The low-$T$ peak and the exponential decay of specific heat below it originate from the thermal activation gap of the triplet excitations out of the gapped singlet-product ground state, while the high-$T$ feature marks the crossover to a regime of nearly uncoupled spins.
Moving into the PS phase at $g=0.7$, the two-peak structure persists, but with notable evolution: the high-$T$ peak becomes sharper, and the low-$T$ peak, signifying the spontaneous transition into the PS phase, shifts to a lower temperature, consistent with the trend reported in both model calculation~\cite{Wietek2019Thermodynamic} and high-pressure specific heat measurement in the material~\cite{Guo2020QuantumPhases,Guo2025Deconfined,Guo_T-linear}. Near the PS–SL boundary ($g=0.77$) and within the SL phase ($g=0.8$), $C$ retains a two-peak shape qualitatively similar to that in the DS phase. 
%
%
Upon entering the AF phase at $g=0.95$, the specific heat evolves into a single-peak structure, characteristic of magnetically ordered systems like that of the square-lattice Heisenberg model~\cite{Chen2018Exponential}.

Fig.~\ref{fig:fig2}(b) shows $C/T$. Dividing by $T$ suppresses the broad high-$T$ shoulder, leaving only peaks at the temperatures corresponding to the low-$T$ features in panel (a). Notably, the lower the temperature of the specific-heat peak, the higher the corresponding peak in $C/T$. For all $g$ values, $C/T$ extrapolates to zero at the lowest temperatures, indicating that $C$ decays faster than linearly in $T$, as shall be seen below, different behavior appears at finite field.

\newsect{Finite field results}
Once we turn on the magnetic field, the phase diagram becomes more complicated and interesting. Start with the DS phase $g=0.6$ in Fig.~\ref{fig:Cv_over_T} (a), one sees that at small $h$ ($h\le 0.6$), the $C/T$ is similar to that at zero-field with a peak around $T\sim 0.1$. However, as $h$ increases, the values of the $C/T$ peak become larger and shifts to a lower temperature since the magnetic field starts to encourage triplet excitations from the dimer singlets by reducing the singlet-triplet gap and these triplet excitations can move in the lattice and introduce density of states (hence the enhanced $C/T$). As their density becomes commensurate with the lattice, the magnetic plateaus will form, as shown in the $m-h$ curve in Fig.~\ref{fig:fig4} (a). Such plateaus have been investigated in previous works~\cite{Kageyama1999Exact,Miyahara1999Exact,mengPhases2008,shiDiscovery2022}, and we observe the 1/3 and 1/2 plateaus in our simulation. 
For magnetic fields above $h=2$, the $m$-$h$ curve rises above the $1/2$ plateau, and the peaks in $C/T$ (Fig.~\ref{fig:Cv_over_T}(a)) shift to even lower temperatures at $h=2$ and $2.5$; we believe they fall below the minimum temperature at which we can collect reliable data. Finally, when $h=3$, the system is fully polarized, and the $C/T$ again shows the gapped behavior within the temperature range we can access.

When increasing $g$ into the PS phase and near the PS-AF phase boundary (at $g=0.7$, $0.77$, and $0.8$), $C/T$ exhibits distinct features, as seen in Fig.~\ref{fig:Cv_over_T}(b)-(d).  At small $h\le 0.6$, $C/T$ also exhibits quick suppression as temperature reduces and extrapolates to zero at $T=0$. Though we failed to generate reliable low-$T$ data at several small $h$ values, e.g., at $g=0.7$ due to limited bond dimension, the peak positions are consistent with the zero field result in Fig.~\ref{fig:fig2}. The peak in $C/T$, which decreases initially at small $h$, exhibits a rebound beginning around $h\approx 1$ accompanied by a shift to higher temperatures. This phenomenon occurs within the interval $h\in[1,1.5]$ in Fig.~\ref{fig:Cv_over_T}(b)-(d), which coincides with the 1/3 plateau region observed in the magnetization curves of Fig.~\ref{fig:fig4}(b)-(d).

However, immediately after the 1/3 plateau regime, $1.5<h<2.2$, very different behavior in $C/T$ is observed. It shows that the $C/T$ will linearly extrapolate to a finite value at $T=0$ (as denoted by the red dashed lines in Fig.~\ref{fig:Cv_over_T} (b), (c), and (d)). The observation of a finite $C/T$ at $T=0$ under a magnetic field aligns remarkably well with the recently discovered DSS reported in experiments~\cite{Guo_T-linear}. For instance, at $h=1.6$ and $g=0.7$, our numerical simulations reveal that $C/T$ becomes temperature-independent (flat) at low temperatures ($T<0.05$). Experimentally, a $T$-linear specific heat is observed under applied pressures of 2.7--3 GPa. Using the linear converting formula $J' = [75 - 8.3P/\text{GPa}] \text{K}$ from Ref.~\cite{Guo2020QuantumPhases}, our simulated temperature $T/J' \sim 0.05$ corresponds to a physical temperature $T \sim 2.5$K. The $T$-linear behavior emerges precisely within the experimental window of 1--2.5K, confirming consistency between our numerical results and the measured data. Our computed $C/T$ data, which follows $C/T = a_0 + a_1 T$ with a finite intercept $a_0$, is compatible with the interpretation of a Dirac spinon state in a magnetic field, featuring a spinon Fermi surface~\cite{shimizuSpin2003,leeU1Gauge2005,motrunichOrbit2006,Ran2007Projected,leeAmperean2007,yamashitaThermodynamic2008,yamashitaGapless2011,Niravkumar2019Magnetic,Wu2025SpinDynamics}.

Once inside the AF phase at $g=0.95$, as shown in Fig.~\ref{fig:Cv_over_T} (e), the $C/T$ as $T\to 0$ goes to zero across the entire field range. At low fields ($h<1$), the behavior resembles the zero-field case with a peak at $T\sim 0.2$ that is at higher temperatures compared to other phases. With increasing field, the peak moves towards lower temperatures at $T\sim 0.1$ and eventually becomes broader when the spins are fully polarized at $h\sim 4$.


\newsect{Discussion}
Overall, our results (the phase diagram in Fig.~\ref{fig:fig1} of the SS model under a magnetic field), 
provide the timely and significant guidance of a novel phase with $T$-linear dependence of the specific heat $C$ at intermediate magnetic field strengths at various couplings. Despite the limitations inherent to our numerical approach (finite system width and bond dimension in the thermal tensor-network methodology), our simulated specific heat data at both zero and finite fields show qualitative and quantitative agreement with previous numerical works~\cite{Wessel2018Thermodynamic,Wietek2019Thermodynamic,Jiménez2021analogue} and recent experimental measurements on SrCu$_2$(BO$_3$)$_2$~\cite{Guo2025Deconfined,Guo_T-linear}. This agreement demonstrates that our calculations successfully capture the essential physics of the material. Most importantly, our theoretical phase diagram is compatible with the rapid experimental developments, pointing to a possible DSS regime in the pressure-magnetic phase diagram of SrCu$_2$(BO$_3$)$_2$~\cite{Guo2020QuantumPhases, Guo2025Deconfined, Guo_T-linear, Zayed20174-spin, Jiménez2021analogue, Cui2023Proximate}. The observation of such a gapless, non-magnetic state emerging from purely magnetic, competing interactions in a Mott insulator model is a significant theoretical result. It underscores the potential for frustrated magnetism, under the interplay of quantum fluctuations and an external field, to liberate deconfined spinons. Our work thus establishes a crucial bridge between recent numerical explorations of spin-liquid physics in the SS model~\cite{Yang2022Quantum, Viteritti2025Transformer, Corboz2025quantum} and its plausible material realization, offering a concrete platform to test emerging theoretical interpretations~\cite{maityEvidence2025, feuerpfeilUnifying2026}.

For future research, more advanced tensor network techniques, such as the tangent-space tensor renormalization group (tanTRG)~\cite{Li2023Tangent} or controlled bond expansion (CBE) methods~\cite{Gleis2022Projector,Gleis2023Controlled,Li2024Time,zhang2025finitet}, could push simulations to larger bond dimensions, enabling a more stringent finite-bond-dimension analysis. Meanwhile, employing two-dimensional algorithms, such as iPEPS adapted for finite temperatures~\cite{Czarnik2019Time,Czarnik2012Projected,Verstraete2004Matrix,Zhang2026Finite}, would allow for a direct study in the thermodynamic limit and verify the robustness and character of the intermediate gapless phase identified in this work. We hope our study stimulates such endeavors to fully unravel the rich physics of the SS model and its fascinating material counterpart SrCu$_2$(BO$_3$)$_2$.

\newsect{Acknowledgment}
We thank Bin-Bin Chen, Min Long, Ting-Tung Wang, and Changkai Zhang for the technical discussion on the thermal tensor-network algorithm. We thank Jing Guo, Shiliang Li and Liling Sun on the collaborations on the experimental exploration of the phase diagram of pressurized and magnetized Shastry-Sutherland Mott insulator SrCu$_2$(BO$_3$)$_2$ over the years~\cite{Guo2020QuantumPhases,Guo2025Deconfined,Guo_T-linear}. We acknowledge the support from the Research Grants Council (RGC) of Hong Kong (Project Nos. 17309822, C7037-22GF, 17302223, 17301924,17301725), the ANR/RGC Joint Research Scheme sponsored by RGC of Hong Kong and French National Research Agency (Project No. A\_HKU703/22). We thank HPC2021 system under the Information Technology Services at the University of Hong Kong~\cite{hpc2021}, as well as the Beijing Paratera Tech Corp., Ltd~\cite{paratera} for providing HPC resources that have contributed to the research results reported within this paper.

\bibliography{bibtex}

@article{Shastry1981Exact,
title = {Exact ground state of a quantum mechanical antiferromagnet},
journal = {Physica B+C},
volume = {108},
number = {1},
pages = {1069-1070},
year = {1981},
issn = {0378-4363},
doi = {https://doi.org/10.1016/0378-4363(81)90838-X},
url = {https://www.sciencedirect.com/science/article/pii/037843638190838X},
author = {B. {Sriram Shastry} and Bill Sutherland}
}

@article{Kageyama1999Exact,
  title = {Exact Dimer Ground State and Quantized Magnetization Plateaus in the Two-Dimensional Spin System ${\mathrm{SrCu}}_{2}({\mathrm{BO}}_{3}){}_{2}$},
  author = {Kageyama, H. and Yoshimura, K. and Stern, R. and Mushnikov, N. V. and Onizuka, K. and Kato, M. and Kosuge, K. and Slichter, C. P. and Goto, T. and Ueda, Y.},
  journal = {Phys. Rev. Lett.},
  volume = {82},
  issue = {15},
  pages = {3168--3171},
  numpages = {0},
  year = {1999},
  month = {Apr},
  publisher = {American Physical Society},
  doi = {10.1103/PhysRevLett.82.3168},
  url = {https://link.aps.org/doi/10.1103/PhysRevLett.82.3168}
}

@article{Miyahara1999Exact,
  title = {Exact Dimer Ground State of the Two Dimensional Heisenberg Spin System ${\mathrm{SrCu}}_{2}({\mathrm{BO}}_{3}){}_{2}$},
  author = {Miyahara, Shin and Ueda, Kazuo},
  journal = {Phys. Rev. Lett.},
  volume = {82},
  issue = {18},
  pages = {3701--3704},
  numpages = {0},
  year = {1999},
  month = {May},
  publisher = {American Physical Society},
  doi = {10.1103/PhysRevLett.82.3701},
  url = {https://link.aps.org/doi/10.1103/PhysRevLett.82.3701}
}

@article{sandvikEvidence2007,
  title = {Evidence for Deconfined Quantum Criticality in a Two-Dimensional Heisenberg Model with Four-Spin Interactions},
  author = {Sandvik, Anders W.},
  journal = {Phys. Rev. Lett.},
  volume = {98},
  issue = {22},
  pages = {227202},
  numpages = {4},
  year = {2007},
  month = {Jun},
  publisher = {American Physical Society},
  doi = {10.1103/PhysRevLett.98.227202},
  url = {https://link.aps.org/doi/10.1103/PhysRevLett.98.227202}
}

@article{maDynamical2018,
  title = {Dynamical signature of fractionalization at a deconfined quantum critical point},
  author = {Ma, Nvsen and Sun, Guang-Yu and You, Yi-Zhuang and Xu, Cenke and Vishwanath, Ashvin and Sandvik, Anders W. and Meng, Zi Yang},
  journal = {Phys. Rev. B},
  volume = {98},
  issue = {17},
  pages = {174421},
  numpages = {12},
  year = {2018},
  month = {Nov},
  publisher = {American Physical Society},
  doi = {10.1103/PhysRevB.98.174421},
  url = {https://link.aps.org/doi/10.1103/PhysRevB.98.174421}
}

@article{Guo2020QuantumPhases,
  title = {Quantum Phases of ${\mathrm{SrCu}}_{2}({\mathrm{BO}}_{3}{)}_{2}$ from High-Pressure Thermodynamics},
  author = {Guo, Jing and Sun, Guangyu and Zhao, Bowen and Wang, Ling and Hong, Wenshan and Sidorov, Vladimir A. and Ma, Nvsen and Wu, Qi and Li, Shiliang and Meng, Zi Yang and Sandvik, Anders W. and Sun, Liling},
  journal = {Phys. Rev. Lett.},
  volume = {124},
  issue = {20},
  pages = {206602},
  numpages = {6},
  year = {2020},
  month = {May},
  publisher = {American Physical Society},
  doi = {10.1103/PhysRevLett.124.206602},
  url = {https://link.aps.org/doi/10.1103/PhysRevLett.124.206602}
}

@article{Guo2025Deconfined,
  author = {Guo, Jing and Wang, Pengyu and Huang, Cheng and Chen, Bin-Bin and Hong, Wenshan and Cai, Shu and Zhao, Jinyu and Han, Jinyu and Chen, Xintian and Zhou, Yazhou and Li, Shiliang and Wu, Qi and Meng, Zi Yang and Sun, Liling},
  title = {Deconfined quantum critical point lost in pressurized SrCu2(BO3)2},
  journal = {Communications Physics},
  year = {2025},
  volume = {8},
  number = {1},
  pages = {75},
  date = {2025-02-22},
  doi = {10.1038/s42005-025-01976-8},
  issn = {2399-3650},
  url = {https://doi.org/10.1038/s42005-025-01976-8}
}

@article{Zayed20174-spin,
  author = {Zayed, M. E. and Rüegg, Ch. and Larrea J., J. and Läuchli, A. M. and Panagopoulos, C. and Saxena, S. S. and Ellerby, M. and McMorrow, D. F. and Strässle, Th. and Klotz, S. and Hamel, G. and Sadykov, R. A. and Pomjakushin, V. and Boehm, M. and Jiménez–Ruiz, M. and Schneidewind, A. and Pomjakushina, E. and Stingaciu, M. and Conder, K. and Rønnow, H. M.},
  title = {4-spin plaquette singlet state in the Shastry–Sutherland compound SrCu2(BO3)2},
  journal = {Nature Physics},
  year = {2017},
  volume = {13},
  number = {10},
  pages = {962--966},
  date = {2017-10-01},
  doi = {10.1038/nphys4190},
  issn = {1745-2481},
  url = {https://doi.org/10.1038/nphys4190}
}

@article{mengPhases2008,
  title = {Phases and magnetization process of an anisotropic Shastry-Sutherland model},
  author = {Meng, Zi Yang and Wessel, Stefan},
  journal = {Phys. Rev. B},
  volume = {78},
  issue = {22},
  pages = {224416},
  numpages = {12},
  year = {2008},
  month = {Dec},
  publisher = {American Physical Society},
  doi = {10.1103/PhysRevB.78.224416},
  url = {https://link.aps.org/doi/10.1103/PhysRevB.78.224416}
}

@article{Cui2023Proximate,
author = {Yi Cui  and Lu Liu  and Huihang Lin  and Kai-Hsin Wu  and Wenshan Hong  and Xuefei Liu  and Cong Li  and Ze Hu  and Ning Xi  and Shiliang Li  and Rong Yu  and Anders W. Sandvik  and Weiqiang Yu },
title = {Proximate deconfined quantum critical point in SrCu<sub>2</sub><b>(</b>BO<sub>3</sub><b>)</b><sub>2</sub>},
journal = {Science},
volume = {380},
number = {6650},
pages = {1179-1184},
year = {2023},
doi = {10.1126/science.adc9487},
URL = {https://www.science.org/doi/abs/10.1126/science.adc9487}
}

@article{Jiménez2021analogue,
  author = {Jiménez, J. Larrea and Crone, S. P. G. and Fogh, E. and Zayed, M. E. and Lortz, R. and Pomjakushina, E. and Conder, K. and Läuchli, A. M. and Weber, L. and Wessel, S. and Honecker, A. and Normand, B. and Rüegg, Ch. and Corboz, P. and Rønnow, H. M. and Mila, F.},
  title = {A quantum magnetic analogue to the critical point of water},
  journal = {Nature},
  year = {2021},
  volume = {592},
  number = {7854},
  pages = {370--375},
  date = {2021-04-01},
  doi = {10.1038/s41586-021-03411-8},
  issn = {1476-4687},
  url = {https://doi.org/10.1038/s41586-021-03411-8}
}

@article{Senthil2004Quantum,
  title = {Quantum criticality beyond the Landau-Ginzburg-Wilson paradigm},
  author = {Senthil, T. and Balents, Leon and Sachdev, Subir and Vishwanath, Ashvin and Fisher, Matthew P. A.},
  journal = {Phys. Rev. B},
  volume = {70},
  issue = {14},
  pages = {144407},
  numpages = {33},
  year = {2004},
  month = {Oct},
  publisher = {American Physical Society},
  doi = {10.1103/PhysRevB.70.144407},
  url = {https://link.aps.org/doi/10.1103/PhysRevB.70.144407}
}

@article{
Senthil2004Deconfined,
author = {T. Senthil  and Ashvin Vishwanath  and Leon Balents  and Subir Sachdev  and Matthew P. A. Fisher },
title = {Deconfined Quantum Critical Points},
journal = {Science},
volume = {303},
number = {5663},
pages = {1490-1494},
year = {2004},
doi = {10.1126/science.1091806},
URL = {https://www.science.org/doi/abs/10.1126/science.1091806}
}

@article{Koga2000Quantum,
  title = {Quantum Phase Transitions in the Shastry-Sutherland Model for ${\mathrm{SrCu}}_{2}({\mathrm{BO}}_{3}{)}_{2}$},
  author = {Koga, Akihisa and Kawakami, Norio},
  journal = {Phys. Rev. Lett.},
  volume = {84},
  issue = {19},
  pages = {4461--4464},
  numpages = {0},
  year = {2000},
  month = {May},
  publisher = {American Physical Society},
  doi = {10.1103/PhysRevLett.84.4461},
  url = {https://link.aps.org/doi/10.1103/PhysRevLett.84.4461}
}

@article{Lee2019Signatures,
  title = {Signatures of a Deconfined Phase Transition on the Shastry-Sutherland Lattice: Applications to Quantum Critical ${\mathrm{SrCu}}_{2}({\mathrm{BO}}_{3}{)}_{2}$},
  author = {Lee, Jong Yeon and You, Yi-Zhuang and Sachdev, Subir and Vishwanath, Ashvin},
  journal = {Phys. Rev. X},
  volume = {9},
  issue = {4},
  pages = {041037},
  numpages = {25},
  year = {2019},
  month = {Nov},
  publisher = {American Physical Society},
  doi = {10.1103/PhysRevX.9.041037},
  url = {https://link.aps.org/doi/10.1103/PhysRevX.9.041037}
}

@article{Corboz2013Tensor,
  title = {Tensor network study of the Shastry-Sutherland model in zero magnetic field},
  author = {Corboz, Philippe and Mila, Fr\'ed\'eric},
  journal = {Phys. Rev. B},
  volume = {87},
  issue = {11},
  pages = {115144},
  numpages = {10},
  year = {2013},
  month = {Mar},
  publisher = {American Physical Society},
  doi = {10.1103/PhysRevB.87.115144},
  url = {https://link.aps.org/doi/10.1103/PhysRevB.87.115144}
}

@article{Nakano2018Third,
author = {Nakano ,Hiroki and Sakai ,T\^{o}ru},
title = {Third Boundary of the Shastry–Sutherland Model by Numerical Diagonalization},
journal = {Journal of the Physical Society of Japan},
volume = {87},
number = {12},
pages = {123702},
year = {2018},
doi = {10.7566/JPSJ.87.123702},
URL = {https://doi.org/10.7566/JPSJ.87.123702}
}

@article{Yang2022Quantum,
  title = {Quantum criticality and spin liquid phase in the Shastry-Sutherland model},
  author = {Yang, Jianwei and Sandvik, Anders W. and Wang, Ling},
  journal = {Phys. Rev. B},
  volume = {105},
  issue = {6},
  pages = {L060409},
  numpages = {7},
  year = {2022},
  month = {Feb},
  publisher = {American Physical Society},
  doi = {10.1103/PhysRevB.105.L060409},
  url = {https://link.aps.org/doi/10.1103/PhysRevB.105.L060409}
}

@article{Viteritti2025Transformer,
  title = {Transformer wave function for two dimensional frustrated magnets: Emergence of a spin-liquid phase in the Shastry-Sutherland model},
  author = {Viteritti, Luciano Loris and Rende, Riccardo and Parola, Alberto and Goldt, Sebastian and Becca, Federico},
  journal = {Phys. Rev. B},
  volume = {111},
  issue = {13},
  pages = {134411},
  numpages = {15},
  year = {2025},
  month = {Apr},
  publisher = {American Physical Society},
  doi = {10.1103/PhysRevB.111.134411},
  url = {https://link.aps.org/doi/10.1103/PhysRevB.111.134411}
}

@misc{Corboz2025Quantum,
      title={Quantum spin liquid phase in the Shastry-Sutherland model revealed by high-precision infinite projected entangled-pair states}, 
      author={Philippe Corboz and Yining Zhang and Boris Ponsioen and Frédéric Mila},
      year={2025},
      eprint={2502.14091},
      archivePrefix={arXiv},
      primaryClass={cond-mat.str-el},
      url={https://arxiv.org/abs/2502.14091}, 
}

@article{Ran2007Projected,
  title = {Projected-Wave-Function Study of the Spin-$1/2$ Heisenberg Model on the Kagom\'e Lattice},
  author = {Ran, Ying and Hermele, Michael and Lee, Patrick A. and Wen, Xiao-Gang},
  journal = {Phys. Rev. Lett.},
  volume = {98},
  issue = {11},
  pages = {117205},
  numpages = {4},
  year = {2007},
  month = {Mar},
  publisher = {American Physical Society},
  doi = {10.1103/PhysRevLett.98.117205},
  url = {https://link.aps.org/doi/10.1103/PhysRevLett.98.117205}
}

@article{Chen2018Exponential,
title = {Exponential Thermal Tensor Network Approach for Quantum Lattice Models},
author = {Chen, Bin-Bin and Chen, Lei and Chen, Ziyu and Li, Wei and Weichselbaum, Andreas},
journal = {Phys. Rev. X},
volume = {8},
issue = {3},
pages = {031082},
numpages = {29},
year = {2018},
month = {Sep},
publisher = {American Physical Society},
doi = {10.1103/PhysRevX.8.031082},
url = {https://link.aps.org/doi/10.1103/PhysRevX.8.031082}
}

@article{Trotter1959,
 ISSN = {00029939, 10886826},
 URL = {http://www.jstor.org/stable/2033649},
 author = {H. F. Trotter},
 journal = {Proceedings of the American Mathematical Society},
 number = {4},
 pages = {545--551},
 publisher = {American Mathematical Society},
 title = {On the Product of Semi-Groups of Operators},
 urldate = {2026-02-03},
 volume = {10},
 year = {1959}
}

@article{Suzuki1990,
title = {Fractal decomposition of exponential operators with applications to many-body theories and Monte Carlo simulations},
journal = {Physics Letters A},
volume = {146},
number = {6},
pages = {319-323},
year = {1990},
issn = {0375-9601},
doi = {https://doi.org/10.1016/0375-9601(90)90962-N},
url = {https://www.sciencedirect.com/science/article/pii/037596019090962N},
author = {Masuo Suzuki}
}

@article{Chen2017Series,
  title = {Series-expansion thermal tensor network approach for quantum lattice models},
  author = {Chen, Bin-Bin and Liu, Yun-Jing and Chen, Ziyu and Li, Wei},
  journal = {Phys. Rev. B},
  volume = {95},
  issue = {16},
  pages = {161104},
  numpages = {5},
  year = {2017},
  month = {Apr},
  publisher = {American Physical Society},
  doi = {10.1103/PhysRevB.95.161104},
  url = {https://link.aps.org/doi/10.1103/PhysRevB.95.161104}
}

@misc{devos2025tensorkit,
      title={TensorKit.jl: A Julia package for large-scale tensor computations, with a hint of category theory}, 
      author={Lukas Devos and Jutho Haegeman},
      year={2025},
      eprint={2508.10076},
      archivePrefix={arXiv},
      primaryClass={cs.MS},
      url={https://arxiv.org/abs/2508.10076}, 
}

@article{Wietek2019Thermodynamic,
  title = {Thermodynamic properties of the Shastry-Sutherland model throughout the dimer-product phase},
  author = {Wietek, Alexander and Corboz, Philippe and Wessel, Stefan and Normand, B. and Mila, Fr\'ed\'eric and Honecker, Andreas},
  journal = {Phys. Rev. Res.},
  volume = {1},
  issue = {3},
  pages = {033038},
  numpages = {20},
  year = {2019},
  month = {Oct},
  publisher = {American Physical Society},
  doi = {10.1103/PhysRevResearch.1.033038},
  url = {https://link.aps.org/doi/10.1103/PhysRevResearch.1.033038}
}

@article{Wessel2018Thermodynamic,
  title = {Thermodynamic properties of the Shastry-Sutherland model from quantum Monte Carlo simulations},
  author = {Wessel, Stefan and Niesen, Ido and Stapmanns, Jonas and Normand, B. and Mila, Fr\'ed\'eric and Corboz, Philippe and Honecker, Andreas},
  journal = {Phys. Rev. B},
  volume = {98},
  issue = {17},
  pages = {174432},
  numpages = {12},
  year = {2018},
  month = {Nov},
  publisher = {American Physical Society},
  doi = {10.1103/PhysRevB.98.174432},
  url = {https://link.aps.org/doi/10.1103/PhysRevB.98.174432}
}

@article{shiDiscovery2022,
author = {Shi, Zhenzhong and Dissanayake, Sachith and Corboz, Philippe and Steinhardt, William and Graf, David and Silevitch, D. M. and Dabkowska, Hanna A. and Rosenbaum, T. F. and Mila, Frédéric and Haravifard, Sara},
year = {2022},
title = {Discovery of quantum phases in the Shastry-Sutherland compound SrCu2(BO3)2 under extreme conditions of field and pressure},
journal = {Nature Communications},
pages = {2301},
volume = {13},
url = {https://doi.org/10.1038/s41467-022-30036-w},
doi = {10.1038/s41467-022-30036-w}
}

@misc{Guo_T-linear,
      title={T-linear specific heat in pressurized and magnetized Shastry-Sutherland Mott insulator SrCu2(BO3)2}, 
      author={Jing Guo and Pengyu Wang and Cheng Huang and Chengkang Zhou and Menghan Song and Xintian Chen and Ting-Tung Wang and Wenshan Hong and Shu Cai and Jinyu Zhao and Jinyu Han and Yazhou Zhou and Qi Wu and Shiliang Li and Zi Yang Meng and Liling Sun},
      year={2026},
      eprint={2602.18229},
      archivePrefix={arXiv},
      primaryClass={cond-mat.str-el},
      url={https://arxiv.org/abs/2602.18229}, 
}

@article{TAKAHASHI1996Thermo,
author = {Takahashi, Yasushi and Umezawa, Hiroomi},
title = {THERMO FIELD DYNAMICS},
journal = {International Journal of Modern Physics B},
volume = {10},
number = {13n14},
pages = {1755-1805},
year = {1996},
doi = {10.1142/S0217979296000817},
URL = {https://doi.org/10.1142/S0217979296000817}
}

@ARTICLE{feuerpfeilUnifying2026,
       author = {{Feuerpfeil}, Andreas and {Shackleton}, Leyna and {Maity}, Atanu and {Thomale}, Ronny and {Sachdev}, Subir and {Iqbal}, Yasir},
        title = "{Unifying Dirac Spin Liquids on Square and Shastry-Sutherland Lattices via Fermionic Deconfined Criticality}",
      journal = {arXiv e-prints},
         year = 2026,
        month = jan,
          eid = {arXiv:2601.19980},
        pages = {arXiv:2601.19980},
          doi = {10.48550/arXiv.2601.19980},
archivePrefix = {arXiv},
       eprint = {2601.19980},
 primaryClass = {cond-mat.str-el},
       adsurl = {https://ui.adsabs.harvard.edu/abs/2026arXiv260119980F}
}

@ARTICLE{maityEvidence2025,
       author = {{Maity}, Atanu and {Ferrari}, Francesco and {Lee}, Jong Yeon and {Potten}, Janik and {M{\"u}ller}, Tobias and {Thomale}, Ronny and {Samajdar}, Rhine and {Iqbal}, Yasir},
        title = "{Evidence for a $\mathbb{Z}_{2}$ Dirac spin liquid in the generalized Shastry-Sutherland model}",
      journal = {arXiv e-prints},
         year = 2024,
        month = dec,
          eid = {arXiv:2501.00096},
        pages = {arXiv:2501.00096},
          doi = {10.48550/arXiv.2501.00096},
archivePrefix = {arXiv},
       eprint = {2501.00096},
 primaryClass = {cond-mat.str-el},
       adsurl = {https://ui.adsabs.harvard.edu/abs/2025arXiv250100096M}
}

@ARTICLE{paratera,
journal={Beijing PARATERA
Tech CO.,Ltd},
url = {https://cloud.paratera.com}
}

@article{hpc2021,
journal={HPC2021, Information Technology Services, The University of Hong Kong},
url={https://hpc.hku.hk/hpc/hpc2021/}
}

@article{shimizuSpin2003,
  title = {Spin Liquid State in an Organic Mott Insulator with a Triangular Lattice},
  author = {Shimizu, Y. and Miyagawa, K. and Kanoda, K. and Maesato, M. and Saito, G.},
  journal = {Phys. Rev. Lett.},
  volume = {91},
  issue = {10},
  pages = {107001},
  numpages = {4},
  year = {2003},
  month = {Sep},
  publisher = {American Physical Society},
  doi = {10.1103/PhysRevLett.91.107001},
  url = {https://link.aps.org/doi/10.1103/PhysRevLett.91.107001}
}

@article{yamashitaThermodynamic2008,
author = {Yamashita, Satoshi and Nakazawa, Yasuhiro and Oguni, Masaharu and Oshima, Yugo and Nojiri, Hiroyuki and Shimizu, Yasuhiro and Miyagawa, Kazuya and Kanoda, Kazushi},
year = {2008},
title = {Thermodynamic properties of a spin-1/2 spin-liquid state in a κ-type organic salt},
journal = {Nature Physics},
pages = {459 - 462},
volume = {4},
url = {https://doi.org/10.1038/nphys942},
doi = {10.1038/nphys942}
}

@article{yamashitaGapless2011,
author = {Yamashita, Satoshi and Yamamoto, Takashi and Nakazawa, Yasuhiro and Tamura, Masafumi and Kato, Reizo},
year = {2011},
title = {Gapless spin liquid of an organic triangular compound evidenced by thermodynamic measurements},
journal = {Nature Communications},
pages = {275},
volume = {2},
url = {https://doi.org/10.1038/ncomms1274},
doi = {10.1038/ncomms1274}
}

@article{leeU1Gauge2005,
  title = {U(1) Gauge Theory of the Hubbard Model: Spin Liquid States and Possible Application to $\ensuremath{\kappa}\mathrm{\text{\ensuremath{-}}}(\mathrm{BEDT}\mathrm{\text{\ensuremath{-}}}\mathrm{TTF}{)}_{2}{\mathrm{Cu}}_{2}(\mathrm{CN}{)}_{3}$},
  author = {Lee, Sung-Sik and Lee, Patrick A.},
  journal = {Phys. Rev. Lett.},
  volume = {95},
  issue = {3},
  pages = {036403},
  numpages = {4},
  year = {2005},
  month = {Jul},
  publisher = {American Physical Society},
  doi = {10.1103/PhysRevLett.95.036403},
  url = {https://link.aps.org/doi/10.1103/PhysRevLett.95.036403}
}

@article{motrunichOrbit2006,
  title = {Orbital magnetic field effects in spin liquid with spinon Fermi sea: Possible application to $\ensuremath{\kappa}\text{\ensuremath{-}}{(\mathrm{ET})}_{2}{\mathrm{Cu}}_{2}{(\mathrm{C}\mathrm{N})}_{3}$},
  author = {Motrunich, Olexei I.},
  journal = {Phys. Rev. B},
  volume = {73},
  issue = {15},
  pages = {155115},
  numpages = {11},
  year = {2006},
  month = {Apr},
  publisher = {American Physical Society},
  doi = {10.1103/PhysRevB.73.155115},
  url = {https://link.aps.org/doi/10.1103/PhysRevB.73.155115}
}

@article{leeAmperean2007,
  title = {Amperean Pairing Instability in the U(1) Spin Liquid State with Fermi Surface and Application to $\ensuremath{\kappa}\mathrm{\text{\ensuremath{-}}}(\mathrm{BEDT}\mathrm{\text{\ensuremath{-}}}\mathrm{TTF}{)}_{2}{\mathrm{Cu}}_{2}(\mathrm{CN}{)}_{3}$},
  author = {Lee, Sung-Sik and Lee, Patrick A. and Senthil, T.},
  journal = {Phys. Rev. Lett.},
  volume = {98},
  issue = {6},
  pages = {067006},
  numpages = {4},
  year = {2007},
  month = {Feb},
  publisher = {American Physical Society},
  doi = {10.1103/PhysRevLett.98.067006},
  url = {https://link.aps.org/doi/10.1103/PhysRevLett.98.067006}
}

@article{Li2023Tangent,
  title = {Tangent Space Approach for Thermal Tensor Network Simulations of the 2D Hubbard Model},
  author = {Li, Qiaoyi and Gao, Yuan and He, Yuan-Yao and Qi, Yang and Chen, Bin-Bin and Li, Wei},
  journal = {Phys. Rev. Lett.},
  volume = {130},
  issue = {22},
  pages = {226502},
  numpages = {8},
  year = {2023},
  month = {Jun},
  publisher = {American Physical Society},
  doi = {10.1103/PhysRevLett.130.226502},
  url = {https://link.aps.org/doi/10.1103/PhysRevLett.130.226502}
}

@misc{zhang2025finitet,
      title={Finite-Temperature Study of the Hubbard Model via Enhanced Exponential Tensor Renormalization Group}, 
      author={Changkai Zhang and Jan von Delft},
      year={2025},
      eprint={2510.25022},
      archivePrefix={arXiv},
      primaryClass={cond-mat.str-el},
      url={https://arxiv.org/abs/2510.25022}, 
}

@article{Gleis2022Projector,
  title = {Projector formalism for kept and discarded spaces of matrix product states},
  author = {Gleis, Andreas and Li, Jheng-Wei and von Delft, Jan},
  journal = {Phys. Rev. B},
  volume = {106},
  issue = {19},
  pages = {195138},
  numpages = {14},
  year = {2022},
  month = {Nov},
  publisher = {American Physical Society},
  doi = {10.1103/PhysRevB.106.195138},
  url = {https://link.aps.org/doi/10.1103/PhysRevB.106.195138}
}

@article{Gleis2023Controlled,
  title = {Controlled Bond Expansion for Density Matrix Renormalization Group Ground State Search at Single-Site Costs},
  author = {Gleis, Andreas and Li, Jheng-Wei and von Delft, Jan},
  journal = {Phys. Rev. Lett.},
  volume = {130},
  issue = {24},
  pages = {246402},
  numpages = {8},
  year = {2023},
  month = {Jun},
  publisher = {American Physical Society},
  doi = {10.1103/PhysRevLett.130.246402},
  url = {https://link.aps.org/doi/10.1103/PhysRevLett.130.246402}
}

@article{Li2024Time,
  title = {Time-Dependent Variational Principle with Controlled Bond Expansion for Matrix Product States},
  author = {Li, Jheng-Wei and Gleis, Andreas and von Delft, Jan},
  journal = {Phys. Rev. Lett.},
  volume = {133},
  issue = {2},
  pages = {026401},
  numpages = {9},
  year = {2024},
  month = {Jul},
  publisher = {American Physical Society},
  doi = {10.1103/PhysRevLett.133.026401},
  url = {https://link.aps.org/doi/10.1103/PhysRevLett.133.026401}
}

@article{Zhang2026Finite,
  title = {Finite temperature dopant-induced spin reorganization explored via tensor networks in the two-dimensional $t\text{\ensuremath{-}}J$ model},
  author = {Zhang, Yintai and Sinha, Aritra and Rams, Marek M. and Dziarmaga, Jacek},
  journal = {Phys. Rev. B},
  volume = {113},
  issue = {8},
  pages = {085113},
  numpages = {13},
  year = {2026},
  month = {Feb},
  publisher = {American Physical Society},
  doi = {10.1103/6pcg-qq4p},
  url = {https://link.aps.org/doi/10.1103/6pcg-qq4p}
}

@article{Verstraete2004Matrix,
  title = {Matrix Product Density Operators: Simulation of Finite-Temperature and Dissipative Systems},
  author = {Verstraete, F. and Garc\'{\i}a-Ripoll, J. J. and Cirac, J. I.},
  journal = {Phys. Rev. Lett.},
  volume = {93},
  issue = {20},
  pages = {207204},
  numpages = {4},
  year = {2004},
  month = {Nov},
  publisher = {American Physical Society},
  doi = {10.1103/PhysRevLett.93.207204},
  url = {https://link.aps.org/doi/10.1103/PhysRevLett.93.207204}
}

@article{Czarnik2012Projected,
  title = {Projected entangled pair states at finite temperature: Imaginary time evolution with ancillas},
  author = {Czarnik, Piotr and Cincio, Lukasz and Dziarmaga, Jacek},
  journal = {Phys. Rev. B},
  volume = {86},
  issue = {24},
  pages = {245101},
  numpages = {6},
  year = {2012},
  month = {Dec},
  publisher = {American Physical Society},
  doi = {10.1103/PhysRevB.86.245101},
  url = {https://link.aps.org/doi/10.1103/PhysRevB.86.245101}
}

@article{Czarnik2019Time,
  title = {Time evolution of an infinite projected entangled pair state: An efficient algorithm},
  author = {Czarnik, Piotr and Dziarmaga, Jacek and Corboz, Philippe},
  journal = {Phys. Rev. B},
  volume = {99},
  issue = {3},
  pages = {035115},
  numpages = {12},
  year = {2019},
  month = {Jan},
  publisher = {American Physical Society},
  doi = {10.1103/PhysRevB.99.035115},
  url = {https://link.aps.org/doi/10.1103/PhysRevB.99.035115}
}

@article{Niravkumar2019Magnetic,
author = {Niravkumar D. Patel  and Nandini Trivedi },
title = {Magnetic field-induced intermediate quantum spin liquid with a spinon Fermi surface},
journal = {Proceedings of the National Academy of Sciences},
volume = {116},
number = {25},
pages = {12199-12203},
year = {2019},
doi = {10.1073/pnas.1821406116},
URL = {https://www.pnas.org/doi/abs/10.1073/pnas.1821406116}}

@article{Wu2025SpinDynamics,
  title = {Spin Dynamics in the Dirac U(1) Spin Liquid ${\mathrm{YbZn}}_{2}{\mathrm{GaO}}_{5}$},
  author = {Wu, Hank C. H. and Pratt, Francis L. and Huddart, Benjamin M. and Chatterjee, Dipranjan and Goddard, Paul A. and Singleton, John and Prabhakaran, D. and Blundell, Stephen J.},
  journal = {Phys. Rev. Lett.},
  volume = {135},
  issue = {4},
  pages = {046704},
  numpages = {8},
  year = {2025},
  month = {Jul},
  publisher = {American Physical Society},
  doi = {10.1103/l93v-f576},
  url = {https://link.aps.org/doi/10.1103/l93v-f576}
}
\bibliographystyle{apsrev4-2}

\newpage
\clearpage
\begin{center}
	\textbf{\large End Matter}
\end{center}
\makeatletter

\setcounter{secnumdepth}{3}
\begin{figure}[htp!]
	\includegraphics[width=\columnwidth]{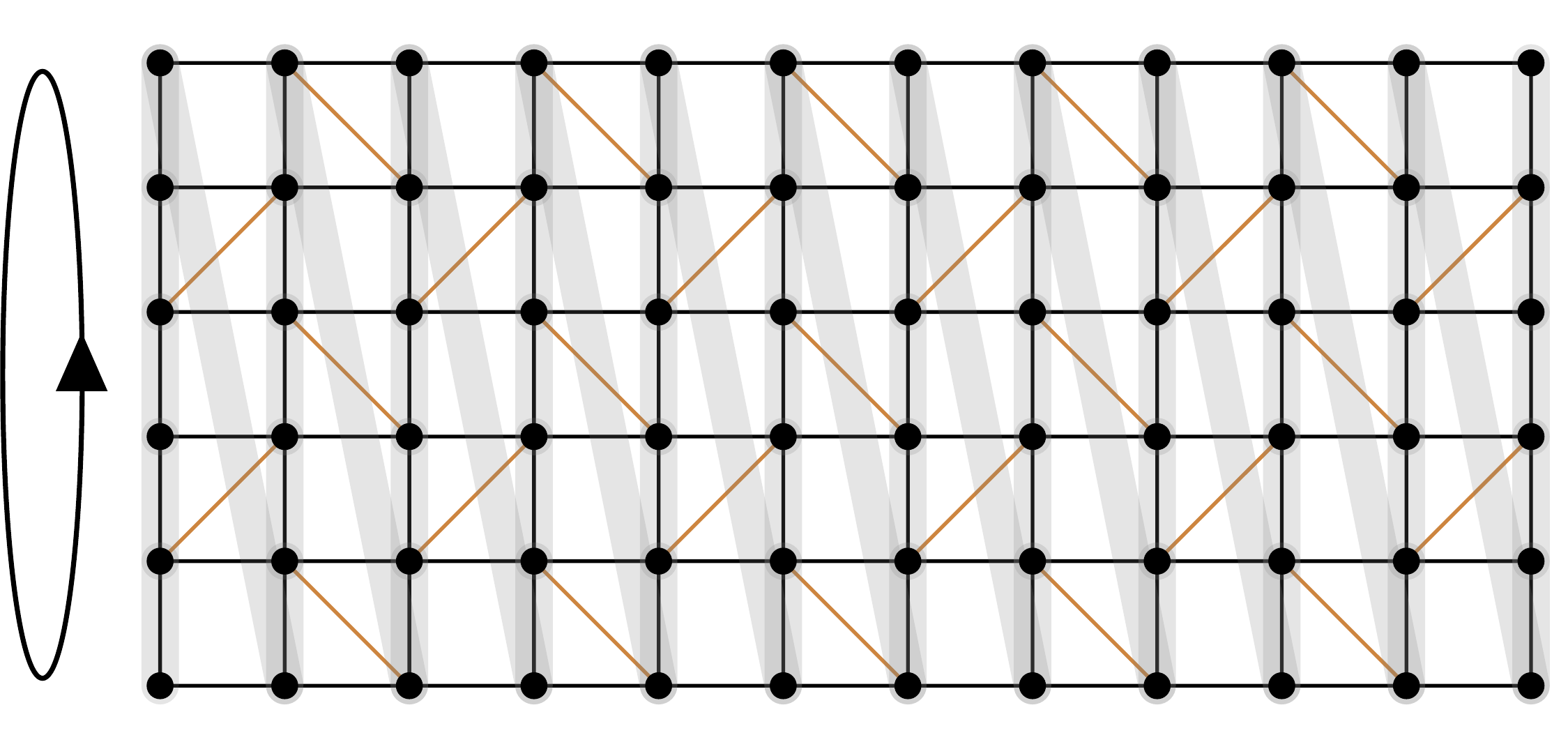}
	\caption{\textbf{Quasi-1D lattice geometry for XTRG simulations.} The cylindrical cluster consists of 3×6×4 sites. Black and brown bonds denote the inter-dimer coupling $J$ and intra-dimer coupling $J'$ from Eq.~\eqref{eq:hamiltonian}, respectively. Periodic boundary conditions are applied along the y-direction. The gray snake-shaped line illustrates the mapping of the cluster onto a one-dimensional chain with long-range interactions.}
	\label{fig:lattice}
\end{figure}

\begin{figure}[htp!]
	\includegraphics[width=\columnwidth]{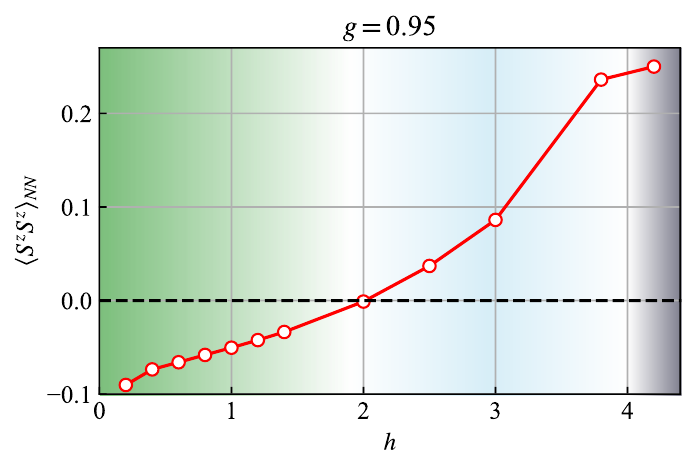}
	\caption{\textbf{Nearest neighbor $\langle S^z S^z\rangle$ at $g=0.95$, in the antiferromagnetic phase} The dashed line marking $\langle S^z S^z\rangle_{NN}=0$ intersects with the red line at $h\approx2$, which we defined as the crossover field strength between AF and AF$'$ (green and blue colored regime). Data is measured at $T\approx 0.01$.}
	\label{fig:SzSz_NN}
\end{figure}

\newsect{XTRG implementation}
The XTRG method begins by constructing a high-temperature density operator $\hat{\rho}(\tau) = e^{-\tau H}$, where $\tau$ represents an infinitesimally small inverse temperature. This initial state can be accurately obtained using techniques such as the Trotter-Suzuki decomposition~\cite{Trotter1959, Suzuki1990} or series-expansion methods~\cite{Chen2017Series}. A key feature of XTRG is its exponential evolution scheme, where the density operator is iteratively squared, i.e., $\hat{\rho}_{n+1} = \hat{\rho}_n \cdot \hat{\rho}_n$, effectively doubling the inverse temperature in each step. This exponential scaling significantly reduces the number of imaginary-time evolution and truncation steps required, and thus enables the accuracy at low temperatures. In practice, we start at a small inverse temperature $\tau_0$ at order of  $\sim 10^{-4}$ by the series expansion method as
$$
\hat\rho_{0} \equiv \hat\rho(\tau_{0}) \simeq \sum_{k=0}^{\mathcal{N}_{\mathrm{c}}} \frac{(-\tau_{0})^{k}}{k !} H^{k}
$$
with the truncated expansion order $\mathcal{N}_{\mathrm{c}}=4$
and perform the doubling of the density matrix 19 times to reach the lowest temperature at order $T\sim 10^{-2}$.

XTRG is an MPO-based thermal tensor network method designed for one-dimensional (1D) systems. To simulate the two-dimensional (2D) Shastry-Sutherland model, we map the 2D cluster onto a snake-like 1D chain with long-range interactions. This approach is illustrated in Fig.~\ref{fig:lattice}, which shows a cluster of width $L_y = 3$ and length $L_x = 6$ in units of the four-site unit cell of the SS lattice; the gray line indicates the snake-like path used to form the 1D chain. We impose periodic boundary conditions along the $y$-direction, resulting in a cylindrical geometry. The results presented in the main text were obtained from simulations on clusters with $L_y = 3$ and $L_x = 12$, corresponding to a cylinder twice as long as that shown in Fig.~\ref{fig:lattice}.

\newsect{Physical observables}
Besides the $C/T$ data shown in Fig.~\ref{fig:Cv_over_T} from which we find a region with $T$-linear dependence of specific heat that might be a possible DSS candidate, Fig.~\ref{fig:fig4} provides other physical observables obtained from XTRG simulation at the lowest temperature $T\approx 0.01$ that help us map out qualitatively the phase diagram in Fig.~\ref{fig:fig1}: thermal entanglement entropy $S_E$ (entanglement entropy of the underlying thermofield double (TFD) state~\cite{Chen2018Exponential,TAKAHASHI1996Thermo}) in the first row, magnetic structure factor for N\'eel order $S^{\pm(z)}(\pi,\pi)$ in the second row and normalized magnetization in the third row. First interesting observation is that the 1/2 plateau in the DS phase at $g=0.6$ coincides with a drastic rise of the $z$-component spin structure factor at $\mathbf{q}=({\pi,\pi})$, which will motivate further theoretical and experimental investigation.

In the PS phase and near the PS-AF boundary (Fig.~\ref{fig:fig4}(b)--(d)), the spin structure factor at wavevector $\mathbf{q}=(\pi,\pi)$, signaling N\'eel-type antiferromagnetic correlations, remains finite at low fields, consistent with earlier zero-field ground state studies~\cite{Yang2022Quantum,Viteritti2025Transformer}. However, upon entering the 1/3 magnetization plateau above $h\sim 1$, $S^{\pm}(\pi,\pi)$ drops sharply and stays small until $h\sim 2$. Concurrently, the thermal entanglement entropy $S_E$, computed at the center of the MPO, exhibits a pronounced dip within the same field interval (upper panels of Fig.~\ref{fig:fig4}(b)--(d)), corresponding directly with the plateau visible in the magnetization curves (lower panels).  

Immediately beyond the plateau, for $g=0.7$, 0.77, and 0.8, the magnetization increases continuously, and a finite $C/T$ extrapolates to zero temperature, indicating a possible DSS regime. This behavior is accompanied by the gradual disappearance of the 1/2 magnetization plateau and a continuous increase in magnetization with increasing magnetic field $h$. Above $h\sim 2$, $S^{\pm}(\pi,\pi)$ shows a slight recovery, most noticeable at $g=0.8$. In analogy with the experimental phase diagram~\cite{Guo_T-linear}, we identify this regime as a field-induced in-plane AF$'$ phase. Finally, the SP phase is most directly identified by its saturation of the normalized magnetization, $M_z/M_{sat} \approx 1$, indicating that the spins are fully polarized. The DS phase is also naturally characterized by the vanishing of both the thermal entanglement entropy $S_E$ and the spin structure factor $S(\pi,\pi)$.

Within the AF phase at $g=0.95$, no transition is observed until the SP phase at $h\sim 4$. To distinguish the low-field AF phase (with both longitudinal and transverse order) from the high-field in-plane AF$'$ phase, we monitor the nearest-neighbor spin correlation along $z$-direction, denoted as $\langle S^zS^z\rangle_{NN}$, which is shown in Fig.~\ref{fig:SzSz_NN}. At low fields, this correlator is negative, consistent with $\langle S^+ S^- \rangle$ due to the rotational symmetry of the zero-field state. As the field increases, spins align with the field, and the correlator turns positive. We define the crossover boundary as the field $h\approx 2$ where $\langle S^z S^z \rangle_{NN}$ is nearly zero, separating the AF and AF$'$ regions.

\end{document}